\documentclass[aps,prl,preprint,showpacs]{revtex4}

\newcommand{\be}{\begin{equation}}
\newcommand{\ee}{\end{equation}}
\newcommand{\bea}{\begin{eqnarray}}
\newcommand{\eea}{\end{eqnarray}}
\renewcommand{\d}{\text{d}}

\usepackage{graphics}
\begin{document}

\title{On the Quantum Density of States and Partitioning an Integer} 

\author{Muoi N. Tran}
\author{M. V. N. Murthy\footnote{Permanent Address: The Institute of 
Mathematical Sciences, Chennai 600 113, India}}
\author{Rajat K. Bhaduri}
\affiliation{Department of Physics and Astronomy, McMaster University,
Hamilton, Ontario, Canada L8S~4M1}

\date{\today}

\begin{abstract}
This paper exploits the connection between the quantum many-particle density 
of states and the partitioning of an integer in number theory. For $N$ bosons 
in a one dimensional harmonic oscillator potential, it is well 
known that the asymptotic ($N\rightarrow\infty$) density of states 
is identical to the Hardy-Ramanujan 
formula for the partitions $p(n)$, of a number $n$ into a sum of integers.  
We show that the same statistical mechanics technique for the density of 
states of bosons in a power-law spectrum yields the partitioning formula for 
$p^s(n)$, the latter being the number of partitions of $n$ 
into a sum of $s^{th}$ powers of a set of integers. By making an appropriate 
modification 
of the statistical technique, we are also able to obtain $d^s(n)$ for 
{\it distinct} partitions. We find that the distinct square 
partitions $d^2(n)$ show pronounced oscillations 
 as a function of $n$ about the smooth curve derived by us. The 
origin of these oscillations from the quantum point of view is discussed.   
After deriving the Erdos-Lehner formula for restricted 
partitions for the $s=1$ case by our method, we generalize it to obtain a 
new formula for distinct restricted partitions. 
\end{abstract}
\vspace{0.5cm}

\pacs{03.65.Sq, 02.10.De, 05.30.-d} 

\maketitle

\section{Introduction}

The N-particle density of states of a self-bound or trapped system has
attracted the attention of physicists for a long time. We have in mind the
work done in nuclear ~\cite{bethe} and particle physics~\cite{hagedorn},
as well as in connection to black-hole entropy~\cite{ashtekar} in recent
times. In nuclear physics, one is generally interested in self-bound
fermions at an excitation energy $E$ that is large compared to the average
single-particle level spacing, but is small compared to the fermi energy
of the nucleus. In this energy range, the density of states is given by
the highly successful Bethe formula~\cite{bethe} that grows as
$exp({a\sqrt E})$, and is insensitive to the details of the
single-particle spectrum. The constant $a$ in the exponent is proportional
to the single particle density of states at the fermi energy $E_F$. In
hadronic physics~\cite{hagedorn}, the many-particle density of states
grows exponentially with $E$, and leads to the concept of a limiting
temperature. The same behavior is found to hold for a bosonic system like
gluons in a bag~\cite{kapusta}. 

It is well known that for ideal bosons in a one-dimensional
harmonic trap, the asymptotic ( $N\rightarrow \infty$ ) density of states 
is the same as the number
of ways of partitioning an integer $n$ into a sum of other integers, and is
given by the famous Hardy-Ramanujan formula~\cite{hardy}. It also grows
exponentially as $\sqrt{E}$, the same as the Bethe formula when $E$ is
identified with $n$. Grossmann and Holthaus~\cite{grossman} have studied
this system, and have used more advanced results from the theory of
partitions~\cite{erdos} to calculate the microcanonical number fluctuation
from the ground state of the system as a function of temperature. Combinatorial methods have also been used to compute the thermodynamic functions for similar systems~\cite{chase}.  In this 
paper we use the N-particle quantum density of states (that may be derived 
using the methods of statistical mechanics) to obtain some novel results   
on the partitioning of an integer into a sum of squares, or a sum of cubes, 
etc. To the best of our knowledge, some of the results pertaining to the  
partitions of an integer to a sum of {\it distinct} powers are new, and will 
be pointed out as they appear in the text. For the harmonic spectrum, we 
are also able to obtain the leading order finite-N 
( Erdos-Lehner~\cite{erdos} ) 
correction to the asymptotic Hardy-Ramanujan formula using our method, and 
then get the corresponding (new) result for distinct restricted partitioning.

In section II of this paper, we consider ideal bosons 
with a single particle spectrum given by a sequence of numbers generated 
by $m^s 
(m=1,2,3,\cdots)$ for a given integer $s\geq 1$. We first derive the
asymptotic many-particle density of states for this system using the
canonical ensemble and in the saddle-point approximation, and show that it
grows exponentially as the $(s+1)$th-root of the excitation energy.
Specifically, in the physically relevant case of a square well, this
result implies that the density of states grows exponentially as the cube
root of energy.  Our general expression for the asymptotic density of states
agrees with the Hardy-Ramanujan formula~\cite{hardy} for $p^s(n)$, 
the number of ways an integer $n$ may be expressed as a sum of $s^{th}$ 
powers of integers. Throughout this paper, we drop the superscript $s$ when 
$s=1$.

The statistical mechanics  method can also be modified to 
obtain asymptotically the number of distinct partitions $d^s(n)$ of an
integer $n$ using the partition function of the fermionic particle spectrum
(excluding the {\it hole} distribution, to be explained later). 
This analysis is presented in 
section III, where the smooth part of the asymptotic density of states 
(which reproduces the average behavior of the distinct partitions) is derived 
using the saddle-point approximation. Interestingly, for the $s=2$ case 
where the integer n is expressed as a sum of distinct squares of integers, 
computations of the exact values of $d^2(n)$ reveal large fluctuation about the smooth average 
curve. These fluctuations wax and wane in a beating pattern. 
The ratio of the amplitude of the oscillations to the smooth part of $d^2(n)$ 
goes to zero as $n\rightarrow\infty$.
From the quantum angle, these oscillations in the many-particle density of 
states have their origin in the fluctuation of the degeneracy of the 
many-particle density of states about the average value. This, in turn, is 
related to the oscillatory part of the single-particle 
density of states of a one-dimensional square-well potential, and the constraint brought about by the Pauli principle.

In section IV, we discuss the corrections to the
saddle-point approximation when the number of particles $N$ is finite. In the
theory of partitions this is known as restricted partitions 
(because of the upper limit on the number of partitions), 
as opposed to the unrestricted partitions discussed in sections II and III.  
We restrict our derivations to the harmonic oscillator ($s=1$) spectrum in 
this section, since the 
canonical partition function is exactly known even for finite $N$ in this case.
For the bosonic case, using this partition function, we derive exponentially  
small corrections for finite $N$, a result that agrees with the 
Erdos-Lehner~\cite{erdos} asymptotic formula for $s=1$. 
This method is then extended for finding the finite-N correction for 
{\it distinct} partitions, a result that to our knowledge is new. 
We conclude with a summary of the main results, as well as some limitations 
of the present work.

\section{The Many-particle density of states}

We first discuss the general statistical mechanical formulation for a  
$N-$particle system. 
The canonical $N-$particle partition function is given by 
\be
Z_N(\beta)=\sum_{E_i^{(N)}}~\eta_i \exp({-\beta{E_i^{(N)}}})=\int_0^{\infty}\rho_N(E) 
\exp(-\beta E) dE~,
\ee
where $\beta$ is the inverse temperature, ${E_i^{(N)}}$ are the eigenenergies 
of the $N-$particle system with degeneracies $\eta_i$, and 
$\rho_N(E)=\sum_i\eta_i\delta(E-{E_i^{(N)}})$ is 
the $N-$particle density of states. 
The density of states $\rho_N(E)$ may therefore be expressed through the 
inverse Laplace transform of the canonical partition function
\be
\rho_N(E) =\frac{1}{2\pi i} \int_{-i\infty}^{i\infty}\exp(\beta E) 
Z_N(\beta) d\beta ~.
\label{rho1}
\ee 
In general, it is not always  possible to do this inversion analytically. 
Note that the single-particle density of states may be decomposed into 
an average (smooth) part, and oscillating components~\cite{brack}. This, in 
turn, results in a smooth part $\overline{\rho}_N(E)$, and an oscillating part 
$\delta \rho_N(E)$~\cite{sakhr} for the $N-$particle density of 
states :
\be
\rho_N(E)=\overline{\rho}_N(E) + \delta\rho_N(E)~.
\label{sad}
\ee
The smooth part $\overline{\rho}_N(E)$ may be obtained by evaluating 
Eq.~(\ref{rho1}) using the saddle-point method~\cite{kubo}. Unlike the 
one-particle case, 
where the oscillating part may be obtained using the periodic orbits in a 
``trace formula''~\cite{brack}, it remains a challenging task to find an 
expression for the oscillating part $\delta\rho_N(E)$~\cite{sakhr}.
In what follows, we shall use the saddle-point method to obtain the smooth 
asymptotic $\overline{\rho}_N(E)$, and identify it with 
the Hardy-Ramanujan formula for $p^s(n)$.  

Before doing this, 
we note that the canonical partition function $Z_N(\beta)$ 
for a set of non-interacting particles with single-particle energies 
$\epsilon_i$, occupancies $\{n_i\}$,  may also be written as 
\be 
Z_N(\beta) = \exp(-\beta E_0^{(N)})
\sum_{\{n_i\}}\Omega(N,E_x) \exp(-\beta E_x\{n_i\}). 
\label{zce}
\ee 
In the above, $E_0^{(N)}$ is the ground-state energy which we set to zero,
and $E_x$ is the excitation energy. The sum is over the allowed occupation
numbers for particles such that $E_x=\sum_i n_i\epsilon_i$. Note that for
a given $E_x$, the number of excited particles in the allowed
configurations may vary from one to a maximum of N. We denote by
$\Omega(N,E_x)$ the total number of such distinct configurations allowed
at an excitation energy $E_x$.  We set the lowest single-particle energy
at zero in order that $E_0^{(N)}=0$, and consider a single-particle spectrum 
$\epsilon_m=m^s$. If now the excitation energy $E_x$
takes only integral values $n$, then $\Omega(N,E)$ is the same as the number
of restricted partitions of $n$, denoted by $p_N^{s}(n)$, and 
asymptotically equivalent to the density of states ${\overline \rho}_N(E)$. 
Omitting the subscript $N$, as in $p^s(n)$, will imply that we are taking 
$N\rightarrow \infty$, corresponding to unrestricted partitioning.

To perform the saddle-point integration of Eq.~(\ref{rho1}), note 
that the integrand may be written as $\exp [S(\beta)]$, where $S(\beta)$ is 
the  entropy given by,
\be
S(\beta) = \beta E + \log Z_N .
\label{entropy}
\ee
Expanding the entropy around the stationary point $\beta_0$ and retaining 
only up to the quadratic term in the expansion in Eq.~(\ref{rho1}) yields
the standard result~\cite{kubo}
\be
\overline{\rho}_N(E) = \frac{exp[S(\beta_0)]}{\sqrt{2\pi S''(\beta_0)}},
\label{rho2}
\ee
where the prime denotes differentiation with respect to inverse 
temperature and 
\be
E=-\left( \frac{\partial \ln Z_N}{\partial \beta}\right)_{\beta_0}. 
\label{saddle}
\ee

We now proceed with a single particle spectrum given by $\epsilon_m =
m^s,$ where the integer $m\geq 1$, and $s>0$ for a system of bosons. The
energy is measured in dimensionless units. For example, when $s=1$ the
spectrum can be mapped on to the spectrum of a one dimensional oscillator
where the energy is measured in units of $\hbar\omega$. For $s=2$,
it is equivalent to setting energy unit as $\hbar^2/2m$, where is m is the
particle mass in a one dimensional square well with unit length. These 
are the
only two physically interesting cases. We however keep $s$ arbitrary even
though for $s >2$ there are no quadratic hamiltonian systems. In
particular $s$ need not even be an integer except to allow a comparison
between the number theoretic results for $p_N^s(n)$ and the density of
states $\rho_N(E)$ that we obtain here. We first obtain the asymptotic
results for unrestricted partitioning by letting $N\rightarrow \infty$ and 
discuss the N-dependent correction 
later. The canonical partition function in this limit may be written as
\be
Z_{\infty}(\beta) = \prod_{m=1}^{\infty}\frac{1}{[1-\exp(-\beta m^s)]},
\label{zprod}  
\ee
where we have used the power-law form for the single particle spectrum. 
By setting $x=\exp(-\beta)$, we see that the bosonic canonical partition 
function is nothing but the generating function \cite{andrews} for 
$p^s(n)$ in number theory, the number of partitions of $n$ into 
perfect $s$-th powers of a set of integers~\cite{hardy} :  
\be
Z_{\infty}(x)=\sum_{n=1}^{\infty} p^s(n) x^n 
=\prod_{n=1}^{\infty}\frac{1}{[1-x^{n^s}]}~. \label{sis}
\ee
In the limit $N\rightarrow \infty$, $p^s(n)$ is the same as $\Omega(E)$ where 
the energy E is replaced by the integer $n$.
In general the above form holds for all $s$ in the limit of 
$N\rightarrow \infty$, but is exact for finite N only for the oscillator 
($s=1$) system \cite{navez,holthaus}. 
Using Eqs.~(\ref{entropy},~\ref{zprod}), and the Euler-MacLaurin series, 
we obtain
\be 
S=\beta E-\sum_{n=1}^{\infty}~\ln [1-\exp(-\beta n^s)] = \beta E 
+ \frac{C{(s)}}{\beta^{1/s}} + {1\over 2}\ln {\beta}-{s\over 2} \ln (2\pi) 
+O(\beta)~,
\label{sinfty}
\ee
where 
\be
C{(s)} = \Gamma(1+\frac{1}{s}) \zeta(1+1/s).
\ee
In the leading order, for determining the stationary point, we ignore the $\ln \beta$ term in the derivatives of $S$ and keeping only the dominant term we obtain
\be 
S^{\prime}(\beta)=E-{1\over s}~ {C{(s)}\over {{\beta^{(1+1/s)}}}}~.
\ee
Therefore the saddle-point is given by  
\be
\beta_0 = \left( \frac{C{(s)}}{s E}\right)^{s/(1+s)}~.
\label{betz}
\ee
The notation may be simplified by setting 
\be
\kappa_s=\left({C{(s)}\over s}\right)^{{s\over {1+s}}}~, 
\ee
so that $\beta_0=\kappa_s~ E^{-{s\over {1+s}}}$.
Substituting this value in the saddle point expression for the density of 
states in Eq.~(\ref{rho2})
\be
\overline{\rho}_{\infty}^{s}(E) = {\kappa_s\over {(2\pi)}^{{(s+1)\over 2}}} 
\sqrt{{s\over {s+1}}} 
E^{-{3s+1\over {2(s+1)}}} \exp\left[\kappa_s (s+1)E^{{1\over {1+s}}}\right]~.
\label{ramanujan}
\label{rhobinfty}
\ee
The RHS of the above equation is {\it identical} to that given for
$p^s(n)$ in \cite{hardy}, the number of ways of expressing $n$ as a sum   
of integers with $s^{th}$ powers, if we replace $E$ by the integer $n$.
For $s=1$, for example, we have
\be
\overline{\rho}_{\infty}(E) = \frac{\exp[\pi\sqrt{\frac{2E}{3}}]}
{4\sqrt{3}E},
\label{rhobk1}
\ee
which is simply the number of partitions of an integer $E$ in
terms of other integers. For example 5=5, 1+4, 2+3, 1+1+3, 1+2+2, 1+1+1+2, 
and 1+1+1+1+1, so $p(5)=7$.
Of course, the above asymptotic formula is not expected to be accurate for 
such a small integer, but it improves in accuracy for large numbers. 

While the "physicists derivation"  of the number partitions has been known
for a while and indeed has been extensively used in the analysis of number
fluctuation in harmonically trapped bose gases \cite{grossman}, the
derivation for a general power law spectrum given above is novel even
though the result was derived long ago by Hardy and Ramanujan \cite{hardy}
using more advanced methods.  Equally interesting from the point of view
of physics is the sensitivity of the bosonic density of states on 
the single-particle spectrum, in contrast to the fermionic Bethe-formula. 
For example, where as in a harmonic well, both fermions and bosons have
the exponential square-root dependence in energy for the density of
states, as given in Eq.~(\ref{rhobk1}), in a square-well only the fermions
obey such a relation when the low temperature expansion is used. For the 
bosonic case, from Eq.~(\ref{ramanujan}), the density of states is
\be
\overline{\rho}_{\infty}^{2}(E) 
=\sqrt{{2\over 3}}{\kappa_2\over {(2\pi)^{3/2}}}
{\exp[3\kappa_2 E^{1/3}]\over E^{7/6}}~.
\label{rhobk2}
\ee
This is the same as the asymptotic formula derived by Hardy and Ramanujan for 
the partition of E into squares, for example $5=1^2+2^2, 1^2+1^2+1^2+1^2+1^2$. 
It is to be noted that in making the identification of $p^{2}(n)$ with 
$\rho^{2}_{\infty}(E)$, $E=n$ is to be identified as the excitation energy 
of the quantum system with a fictitious ground state at zero energy added to 
the square well.

In Fig.~(\ref{fig1}) we show a comparison between the exact (computed) 
$p(n)$ (continuous line), and $\overline{\rho}_{\infty}(n)$ 
(dashed line), as given by Eq.~(\ref{rhobk1}). We note that the 
Hardy-Ramanujan formula works 
well even for small $n$. Similarly, in Fig.~(\ref{fig2}), the computed 
$p^2(n)$ is compared with $\overline{\rho}^{2}_{\infty}(n)$, as given by 
Eq.~(\ref{rhobk2}). It will be noted from Fig.~(2) that the computed 
$p^2(n)$ has step-like discontinuities, unlike the smooth behavior of 
$\overline{\rho}^{2}_{\infty}(n)$, specially for small $n$. We should 
remind the reader that these results are not new, and the corrections to 
the leading order Hardy-Ramanujan formula are also known in the number theory 
literature. We shall, however, obtain some new results using our method 
for distinct partitions $d^2(n)$ in the next section.     

Before we conclude this section, we note that keeping terms of order
$\beta$ in the saddle point expansion of $S$ merely shifts the energy E by
the coefficient of the term proportional to $\beta$. For the $s=1$ case 
in Eq.~(\ref{sinfty}), there is indeed a term like $-{1\over {24}} \beta$, 
leading to the replacement of $E$ by $(E-{1\over {24}})$ in Eq.~(\ref{rhobk1}). 
The resulting asymptotic
expression for the density is the first term of the exact convergent
series for partitions obtained by Rademacher \cite{rad}. 
Interestingly, for $s=2$, there is no term of order $\beta$ in the
Euler-Maclaurin expansion. A similar situation prevails for distinct 
partitions as will be shown in the next section.

\section{Asymptotic density of states with distinct partitions}

We now modify the method to obtain distinct partitions of an 
integer $n$ into $s^{th}$ powers, to be denoted by $d^s(n)$. 
For example, for $s=1$, $n= 5$, the number of distinct
integer partitions are 5, 2+3, and 1+4, so $d(5)=3$. 
For distinct partitions, the first guess would be to use the 
fermionic partition function instead of 
the bosonic one of the previous section since 
distinctiveness of the parts is immediately ensured by the Pauli principle.  
However, there is a problem here 
which we illustrate using the $s=1$ spectrum. For this case 
the fermionic partition function of non-interacting particles is 
given by (setting $x=\exp(-\beta)$ as before),
\be 
Z_N(x)= x^{N^2/2} \prod_{m=1}^{N} \frac{1}{(1-x^m)} = 
x^{N^2/2}\sum_{n=0}^{\infty} \Omega(N,n) x^n~, 
\ee
which is the same as the bosonic partition function in a harmonic potential, 
except for the prefactor which is related to the ground state energy of
$N$ particles in the trap. Obviously, the $\Omega(N,n)$ is the same for
both fermions and bosons even though $d_N(n)$ is different from $p_N(n)$. 
This is because the quantum mechanical ground state of
fermions consists of occupied levels up to the fermi energy, unlike the bosons
which all occupy the lowest energy state. Thus, for the fermions at any 
excitation energy, one should consider the distribution among particles as 
well as holes, each of which is separately distinct \cite{muoi}, and  
obey the Pauli principle. As we show below, the particle distribution at 
a given excitation energy measured from the Fermi energy identically 
reproduces (the unrestricted) but distinct partitions of an integer $n$,  
when $n$ is identified with the excitation energy.  

The relevant "partition" function for the $m^s$ spectrum is given by,
\be 
\ln Z_{\infty}(\beta) = \sum_{m=1}^{\infty} \ln [1+\exp(-\beta m^s)]~,
\label{power}
\ee
and the entropy $S(\beta)$ is obtained as usual by adding $\beta E$ to 
the above expression. Notice that this resembles the entropy of an N-fermion 
system, but with the chemical potential $\mu=0$. In the normal
N-fermion system at any given excitation energy the number of macro states
available depends on the distribution of both particles above the Fermi
energy and holes below the Fermi energy in the ground states. By setting
$\mu=0$ we are ignoring the hole distribution but only taking into account
the states associated with the particle distribution. Because of Pauli
principle implied in the above form for the entropy, only distinct
partition of energy E is allowed. Again, using the variable $x=\exp(-\beta)$ 
in Eq.~(\ref{power}), $Z_{\infty}(x)$ above is seen to be the generating 
function for distinct partitions $d^s(n)$ of an integer $n$ into 
$s^{th}$ powers of other integers~\cite{andrews}.  

Once this point is noted, the rest of the calculation proceeds as in the
case of bosons and we obtain the following expression using the
Euler-MacLaurin series
\be 
S(\beta) = \beta E 
+ \frac{D{(s)}}{\beta^{1/s}} -\frac{1}{2}\ln(2) + O(\beta)~,
\label{sferm} 
\ee
where 
\be
D{(s)} = \Gamma(1+\frac{1}{s}) \eta(1+1/s),
\ee
where $\eta(s)=\sum_{l=1}^{\infty} {(-1)^{l-1}\over {l^s}}$ denotes the 
alternating zeta function. Note there is no $\log(\beta)$ term in 
Eq.~(\ref{sferm}). The saddle point $\beta_0$ is obtained 
by setting $S^{\prime}(\beta_0)=0$ as before. Defining 
\be
\lambda_s = (D{(s)}/s)^{s/(s+1)}
\ee
and using Eq.~(\ref{rho2}), 
we obtain
\be
\overline{\rho}^{s}_{\infty(F)}(E) = \sqrt{s\lambda_s}~{{exp\left[(1+s)\lambda_s 
E^{1\over {1+s}}\right]\over {2\sqrt{\pi(1+s)E^{2s+1\over {s+1}}}}}}~,
\label{rhofinfty}
\ee
where the subscript $(F)$ in $\overline{\rho}$ is the remind the reader that Fermi statistics has been used (with $\mu=0$).
Once again for s=1 we recover the well known asymptotic formula for the 
unrestricted 
but distinct partitions $d(n)$ of an integer \cite{abramowitz}, namely
\be
\overline{\rho}_{\infty(F)}(E) = \frac{\exp[\pi\sqrt{\frac{E}{3}}]}
{4\times3^{1/4}E^{3/4}}~,
\label{rhofk1}
\ee
where, as usual, $E$ should be read as $n$. Similarly the asymptotic 
expression for $d^s(n)$ is given by in Eq.~(\ref{rhofinfty}). We have not found 
this general expression in the literature, though we believe it must be known. 
As we shall presently see, however, a remarkable finding is made when this 
asymptotic formula is compared with the exact computation of $d^s(n)$ for 
$s=2$. First, however, in Fig.~(\ref{fig3}), we show a comparison of the 
asymptotic density $\overline{\rho}_{\infty(F)}$ and the exact distinct partitions $d(n)$ of integer $n$ for $s=1$. As in the 
case of bosonic partitions $p(n)$, the asymptotic formula for $d(n)$ 
works reasonably, except for $n<10$. But the really interesting result 
is shown in Fig.~(\ref{fig4}) where we compare Eq.~(\ref{rhofinfty}) for  
$s=2$ with exact computations of $d^2(n)$. The asymptotic density of states 
follows the average of the exact $d^2(n)$ closely, but there are pronounced 
beat-like structure superposed on this smooth curve. This has come about 
because we have joined the computed points of $d^2(n)$ for discrete $n$'s 
by zig-zag lines.   
Note that compared to $d(n)$, the magnitude of $d^2(n)$ is very small, and 
this is one reason that the fluctuations in $d^2(n)$ look so prominent. We have checked 
numerically, however, that the ratio of the amplitude of the oscillations 
to its smooth average value decreases from about 1.5 to 0.2 as $n$ is 
increased to 1000. This means that for $n\rightarrow \infty$, the smooth 
part will eventually mask the fluctuations.

Although we cannot analytically reproduce these fluctuations in the 
many-particle 
density of states (or equivalently in $d^2(n)$), we can show from the 
quantum point of view that the smooth part $\overline{\rho}^2_{\infty(F)}$ arises strictly from the smooth part of the single-particle density of states.
 To make this 
point, let us derive the single-particle density of states, $g(\epsilon)$, 
for the $n^2$ spectrum.  
We begin with the knowledge of the exact single-particle spectrum, and write 
the canonical partition function:
\be
Z_1(\beta)=\sum_{n=1}^\infty \exp (-\beta  n^2)~.
\ee
To express this in a tractable form for Laplace-inverting, we use the 
(exact) Poisson sum formula
\be
\sum_{n=-\infty}^\infty F(n)= \sum_{q=-\infty}^\infty {\cal F}(q)~,
\ee
where 
\be
{\cal F}(q)= \int_{-\infty}^\infty dn~ F(n) \exp(2\pi iqn)~.
\ee 
Taking $F(n)=\exp(-\beta  n^2)$ then gives ${\cal F}(q)=
\sqrt{\pi/(\beta )} \exp(-\pi^2 q^2/(\beta ))$.
Using this result, we obtain 
\be
Z_1(\beta)= {1\over 2}~\left({\pi\over {\beta }}\right)^{1/2} -{1\over 2} 
+\left({\pi\over {\beta }}\right)^{1/2}~\sum_{q=1}^\infty \exp
(-\pi^2 q^2/(\beta))~.
\label{zlin}
\ee
On Laplace-inverting term by term, we obtain the exact result for the 
single-particle density of states :
\bea
g(\epsilon)&=&{1\over {2 \sqrt{\epsilon }}}- {1\over 2} \delta (\epsilon) +
{1\over {\sqrt{\epsilon}}} \sum_{q=1}^{\infty} {\rm cos\left(2\pi q \sqrt{
\epsilon}\right)}~,\label{kya}\\
           &=&\overline{g}(\epsilon)+\delta g(\epsilon)~,
\label{kyabat}
\eea
where $\overline{g}(\epsilon)$ is the ``smooth'' part consisting of the 
first two terms on the RHS of Eq.~(\ref{kya}), and $\delta g(\epsilon)$ denotes 
the remaining oscillating terms. 
We can now evaluate Eq.~(\ref{power}) for $s=2$ using the above $g(\epsilon)$ :
\be
\ln Z_{\infty} = \int_0^{\infty} g(\epsilon) \ln [1+\exp(-\beta\epsilon )]~
d\epsilon~.
\ee
Evaluating the integrals, and adding $\beta E$ to it, we get the entropy 
\be
S(\beta)= \beta E 
+ \frac{D{(2)}}{\beta^{1/2}} -\frac{1}{2}\ln(2)  + 
{\sqrt{\pi\over \beta}}\sum_{q=1}^{\infty}~\sum_{l=1}^{\infty}~{(-)^{l+1}\over 
l^{3/2}} \exp\left({-\pi^2 q^2\over {\beta l}}\right)~.
\label{osc}
\ee
We note that the first two terms on the RHS of the above equation are the 
same as obtained earlier in Eq.~(\ref{sferm}) using the Euler-MacLaurin 
expansion. These yielded the smooth many-body density of states given by 
Eq.~(\ref{rhofinfty}) on using the saddle-point approximation. The term with the double sum in Eq.~(\ref{osc}), which arise from $\delta g(\epsilon)$ in 
Eq.~(\ref{kyabat}) and Fermi statistics, must be the source of the fluctuations seen in 
the density of states in Fig.~4 (the same $\delta g(\epsilon)$, when used in the bosonic case, gives a very different contribution to $S(\beta)$). In principle, exact Laplace inversion 
of $\exp [S(\beta)]$, where $S(\beta)$ is given by Eq.~(\ref{osc}), should yield  
the fluctuating degeneracies of the quantum states with $E$, and hence of 
$d^2(n)$. We have not 
been able, however, to do this Laplace inversion. Since the oscillation in 
the exact partitions $d^2(n)$ resemble a beat-like structure, at least two 
frequencies must be interfering to give the pattern. Further work is  
needed to unravel this interesting point.

\section{Finite size corrections, or restricted partitions}

The smooth part of the many-particle density of states was derived in the 
previous sections for a
system with $N\rightarrow \infty$, that corresponded to unrestricted  
partitions. We now apply the same method to obtain the asymptotic density of 
states for systems with finite size, that is when the number of particles is 
kept 
finite and equal to $N$. This corresponds to allowing the number of parts to 
be at most $N$. Consider, for example, for $s=1$, $N=4$, $n=5$. Then, in 
restricted partitioning, the allowed partitions are 5, 4+1, 3+2, 3+1+1, 
2+2+1, and 2+1+1+1. The partition with 5 parts, 1+1+1+1+1 is not allowed, 
since the number of parts in this case is greater than 4. The above example 
is for restricted case that includes identical parts in a partition. These 
will be denoted by $p_N^s(n)$ in general, but for $s=1$, the superscript 
will be dropped as usual. For the above example with restricted and distinct  
partitions, however, only 5, 1+4, and 2+3 are allowed. We denote such 
partitioning by $d_N^s(n)$ in general. 
In this section, we restrict to $s=1$, and first present the leading order 
asymptotic expression for $p_N(n)$, 
using our method of calculating $\overline{\rho}_N(E)$. This  
result is already known in the literature by the Erdos-Lehner 
formula~\cite{erdos}, but is derived here because we generalize it for 
obtaining the asymptotic 
expression for $d_N(n)$. To the best of our knowledge, this is a new result. 

\subsection{Asymptotic formula for $p_N(n)$}

The $N-$boson canonical partition function in this case is exactly known :
\be 
\ln Z_N (\beta) = -\sum_{m=1}^{N}\ln[1-\exp(-\beta m)].  
\label{znce} 
\ee
The canonical entropy $S_N$ is obtained as before by adding 
$\beta E$ to the above equation.
Expanding the above using Euler MacLaurin series, 
and assuming that $N$ is large so that $x=exp(-\beta N) <<1$, even though 
$\beta << 1$. We then obtain 
\be 
S_N(\beta) = S_{\infty}(\beta) -{\exp(-\beta N)}[\frac{1}{\beta}-{1\over 2}], 
\label{sn3}
\ee
The stationary point is determined as before by the condition in
Eq.~(\ref{saddle}) and for N large it is the same as in Eq.~(\ref{betz}). 
Substituting this in the saddle point expression for the density of 
states in Eq.~(\ref{rho2}) we get 
\be
\overline{\rho}_{N}(E) = \overline{\rho}_{\infty}(E) \exp\left[ 
-\left(\frac{\sqrt{6E}}{\pi }-{1\over 2}\right)\exp\left(-\frac{\pi 
N}{\sqrt{6E}}\right)\right]. 
\label{rhobn1}
\ee
The above expression reproduces the well known correction to the 
unrestricted partitions due to the restriction on the number of particles 
(see Erdos and Lehner \cite{hardy}) apart from the 
constant term proportional to $1\over 2$ in the exponent.  This constant 
may, however, be neglected when $E$ is large. Using the conditions $\beta_0 << 1$ and $\beta_0N >> 1$, we see that formula (\ref{rhobn1}) is valid in the region $C(1) << E << C(1)N^2$, where $C(1) \approx 1.645$.  In Fig.~5 we compare the two differences, $\left[\overline{\rho}_{\infty}(E)-p_N(n)\right]$, and $\left[\overline{\rho}_N(E)-p_N(n)\right]$  for $N=20$ (Fig.~5a), and $N=30$ (Fig.~5b).  In the above, $\overline{\rho}_{\infty}(E)$ is obtained from Eq.~(\ref{rhobk1}), $\overline{\rho}_{20}(E)$ is the Erdos and Lehner formula as given by Eq.~(\ref{rhobn1}), and $p_{20}(n)$ is the exact (computed) restricted partitions. Clearly, the former is much larger than the latter, indicating that Eq.~(\ref{rhobn1}) gives a better approximation to the exact values for restricted partitions. 

\subsection{Asymptotic formula for $d_N(n)$}
Next we present the finding of an equivalent asymptotic formula to Eq.~(\ref{rhobn1}) for the restricted and distinct partition.  This brings us back to the fermionic particle spectrum as discussed in section II and Ref.~\cite{muoi}.  Eq.~(\ref{power}) of section II does not apply here, however, since it is applicable only for the unrestricted distinct partition, ie: $N \rightarrow \infty$.  What we need is the exact canonical partition function for the particle space.  From number theory \cite{rademacher,muoi}, we found a formula for the (exact) number of ways of partitioning an integer $n$ to at most N distinct parts:
\be
d_N(n)=\sum_i^Np_i(n-\frac{i(i+1)}{2}),
\label{distinctO}
\ee
where $p_i(n)$ is the (exact) number of partitions of $n$ to at most $i$ parts, which may be generated by the partition function given by Eq.~(\ref{znce}).  Eq.~(\ref{distinctO}) implies that the partition or generating function for the restricted and distinct partition is given by:
\bea
Z^{(d)}_N (\beta) &=& \sum_{i=1}^N x^{i(i+1)/2}\prod_{n=1}^i\frac{1}{\left(1-x^n\right)}, ~\nonumber \\ 
                  &=& \prod_{n=1}^{\infty}\left(1+x^n\right) -\sum_{i=N+1}^{\infty} x^{i(i+1)/2}\prod_{n=1}^i\frac{1}{\left(1-x^n\right)}.
\label{distinctZ}
\eea
The first term on the right hand side of Eq.~(\ref{distinctZ}) is the generating function for the unrestricted distinct partition Eq.~(\ref{power}), and the second term is a sum of the generating functions for the restricted non-distinct partition Eq.~(\ref{znce}) with the integer shifted to $i(i+1)/2$.  To find an asymptotic formula for the restricted distinct partition $d_N(n)$, as usual, we take inverse Laplace transform of Eq.~(\ref{distinctZ}):
\bea
d_N(n) &=& L^{-1}_\beta \left\{\prod_{n=1}^{\infty}\left(1+x^n\right)\right\}-\sum_{i=N+1}^{\infty}L^{-1}_\beta \left\{x^{\triangle}\prod_{n=1}^i\frac{1}{\left(1-x^n\right)}\right\}~, \nonumber \\
              &=& d(n)- \sum_{i=N+1}^{\infty}p_{i}(n-\triangle)~, \nonumber \\
              &\sim& \overline{\rho}_{\infty(F)}(E) - \sum_{i=N+1}^{\infty}\overline{\rho}_i(E-\triangle)~,\nonumber  \\ 
              &=& \frac{\exp[\pi\sqrt{\frac{E}{3}}]}{4\times3^{1/4}E^{3/4}} -\sum_{i=N+1}^{\infty} \overline{\rho}_{\infty}(E-\triangle) \exp\left[-\left(\frac{\sqrt{6(E-\triangle)}}{\pi }-{1\over 2}\right)\exp\left(-\frac{\pi N}{\sqrt{6(E-\triangle)}}\right)\right]~, \nonumber \\  
              &=& \overline{\rho}_{N(F)}(E)~.
\label{dn}
\eea
where $\triangle \equiv i(i+1)/2$, $x \equiv exp(-\beta)$ and $n$ is identified with $E$.  Note that since the asymptotic expression for the restricted partition $\overline{\rho}_{N}(E)$ is valid only for $C(1) << E << C(1)N^2$, Eq.~(\ref{dn}) is thus valid only in this range.  Fig.~6 displays the two differences, $\left[\overline{\rho}_{\infty(F)}(E)-d_{N}(n)\right]$, and $\left[\overline{\rho}_{N(F)}(E)-\d_N(n)\right]$ for $N=20$ (Fig.~6a), and $N=30$ (Fig.~6b).  In the above differences, $\overline{\rho}_{\infty(F)}(E)$ is obtained from Eq.~(\ref{rhofk1}), $\overline{\rho}_{N(F)} (E)$ from Eq.~(\ref{dn}), and  $d_{N}(n)$ is the exact (computed) restricted distinct partitions.  Again, similar to the non-distinct case (Fig.~5), the N-correction asymptotic formula gives a better approximation to the exact finite N partition than the infinite distinct one. 

\section{Discussion}
This work emphasizes the connection between the many-body quantum density 
of states in a power-law spectrum with the number theoretic partitions 
$p^s(n)$, and the distinct partitions $d^s(n)$. This was already well-known 
to the physics community for $p(n)$. Most of the results derived in this paper 
are known in the mathematical literature, with the possible exception of 
the the asymptotic formula for $d^s(n)$ (Eq.~(\ref{rhofinfty})), and the  
generalized formula (\ref{dn}) for restricted distinct partitions. 
Perhaps the most interesting result in the paper is shown in Fig.~4, where 
the fluctuations in $d^2(n)$ are shown. These oscillations in this number-theoretic quantity are linked, from the quantum mechanical point of view, 
to the oscillating part of the density of 
states in a square-well potential, and the Pauli principle. However, we are not able to completely 
demonstrate this point to our satisfaction because of the difficulty of 
Laplace inversion of exponentiated quantities. This is left for the future.

This work was supported by NSERC (Canada). We would like to thank Ranjan 
Bhaduri for many helpful hints and discussions as well as pointing out
appropriate references. Thanks are also due to Jamal Sakhr and Oliver King 
for their advice. MVN would like to thank K. Srinivas for discussions and the Department of Physics and Astronomy, McMaster University for hospitality.

\newpage

\begin{figure}[h]
\caption{Comparison of the exact $p(n)$ (solid line) and the 
asymptotic $\overline{\rho}_{\infty}(E)$ (dashed line), obtained from Eq.~(\ref{rhobk1}) for $s=1$.} 
\label{fig1}
\end{figure}

\begin{figure}[h]
\caption{Comparison of the exact $p^2(n)$ (solid line) and the 
asymptotic $\overline{\rho}^2_{\infty}(E)$ (dashed line), obtained from Eq.~(\ref{rhobk2}) for $s=2$.}
\label{fig2} 
\end{figure}

\begin{figure}[h]
\caption{Comparison of the exact $d(n)$ (solid line) and the asymptotic $\overline{\rho}_{\infty(F)}(E)$ (dashed line), obtained from Eq.~(\ref{rhofk1}) for $s=1$ and distinct partitions.}
\label{fig3} 
\end{figure}

\begin{figure}[h]
\caption{Comparison of the exact $d^2(n)$ (solid line) and the asymptotic $\overline{\rho}^2_{\infty(F)}(E)$ (dashed line), obtained from Eq.~(\ref{rhofinfty}) for $s=2$ and distinct partitions.  Note that the y-axis is no longer in log scale.}
\label{fig4} 
\end{figure}

\begin{figure}[h]
\caption{(a) Comparison of $\left[\overline{\rho}_{\infty}(E)-p_{20}(n)\right]$ (dotted line) and $\left[\overline{\rho}_{20}(E)-p_{20}(n)\right]$ (solid line) for $N=20$, where $\overline{\rho}_{\infty}(E)$ is obtained from Eq.~(\ref{rhobk1}), $\overline{\rho}_{20}(E)$ is the Erdos and Lehner formula as given by Eq.~(\ref{rhobn1}), and $p_{20}(n)$ is the exact (computed) restricted partitions. (b) Same for $N=30$.}
\label{fig5} 
\end{figure}

\begin{figure}[h]
\caption{(a) Comparison of $\left[\overline{\rho}_{\infty(F)}(E)-d_{20}(n)\right]$ (dotted line) and $\left[\overline{\rho}_{20(F)}(E)-d_{20}(n)\right]$ (solid line) for $N=20$, where $\overline{\rho}_{\infty(F)}(E)$ is obtained from Eq.~(\ref{rhofk1}), $\overline{\rho}_{20(F)}(E)$ from Eq.~(\ref{dn}), and  $d_{20}(n)$ is the exact (computed) restricted distinct partitions. (b) Same for $N=30$.}
\label{fig6} 
\end{figure}

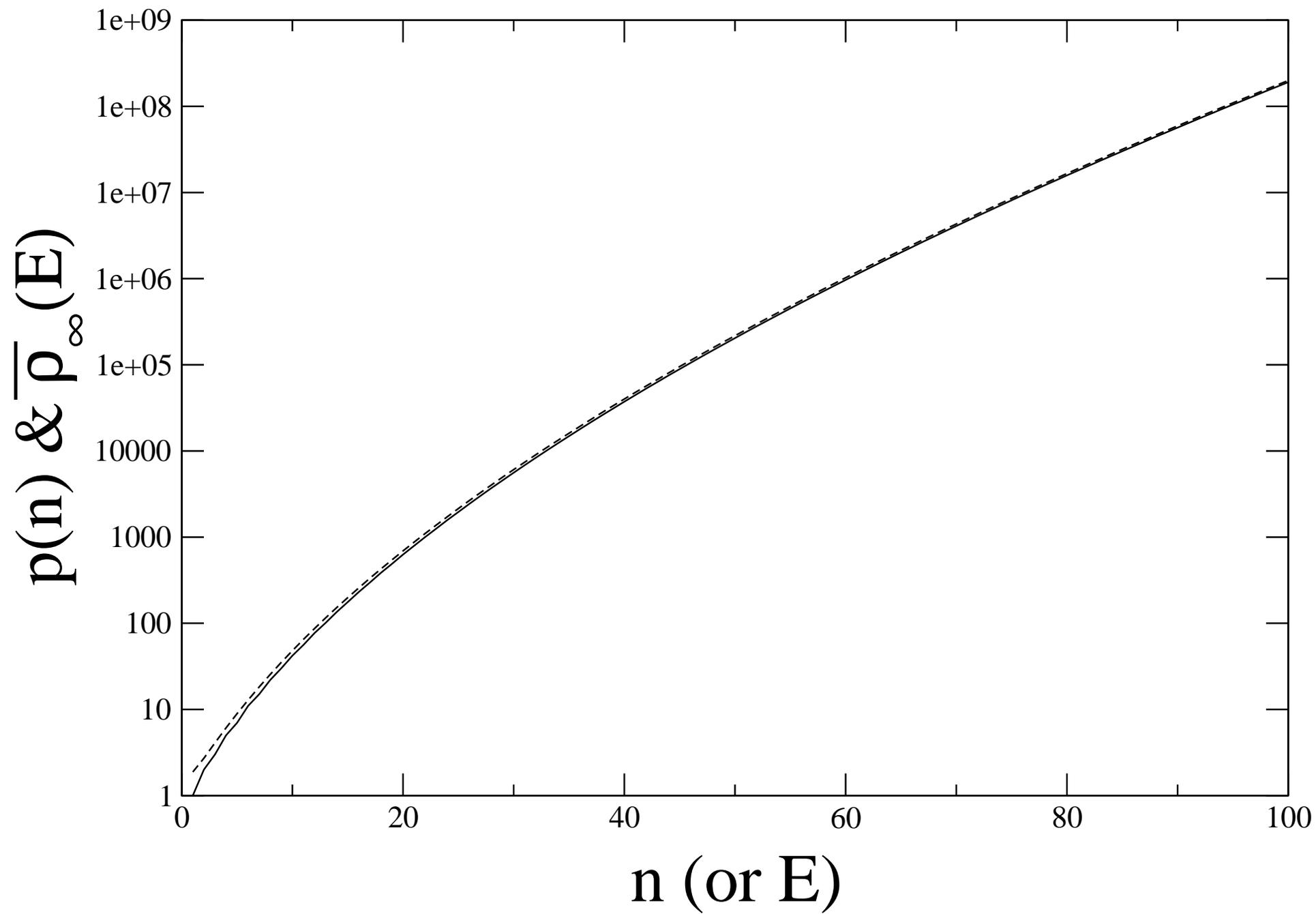
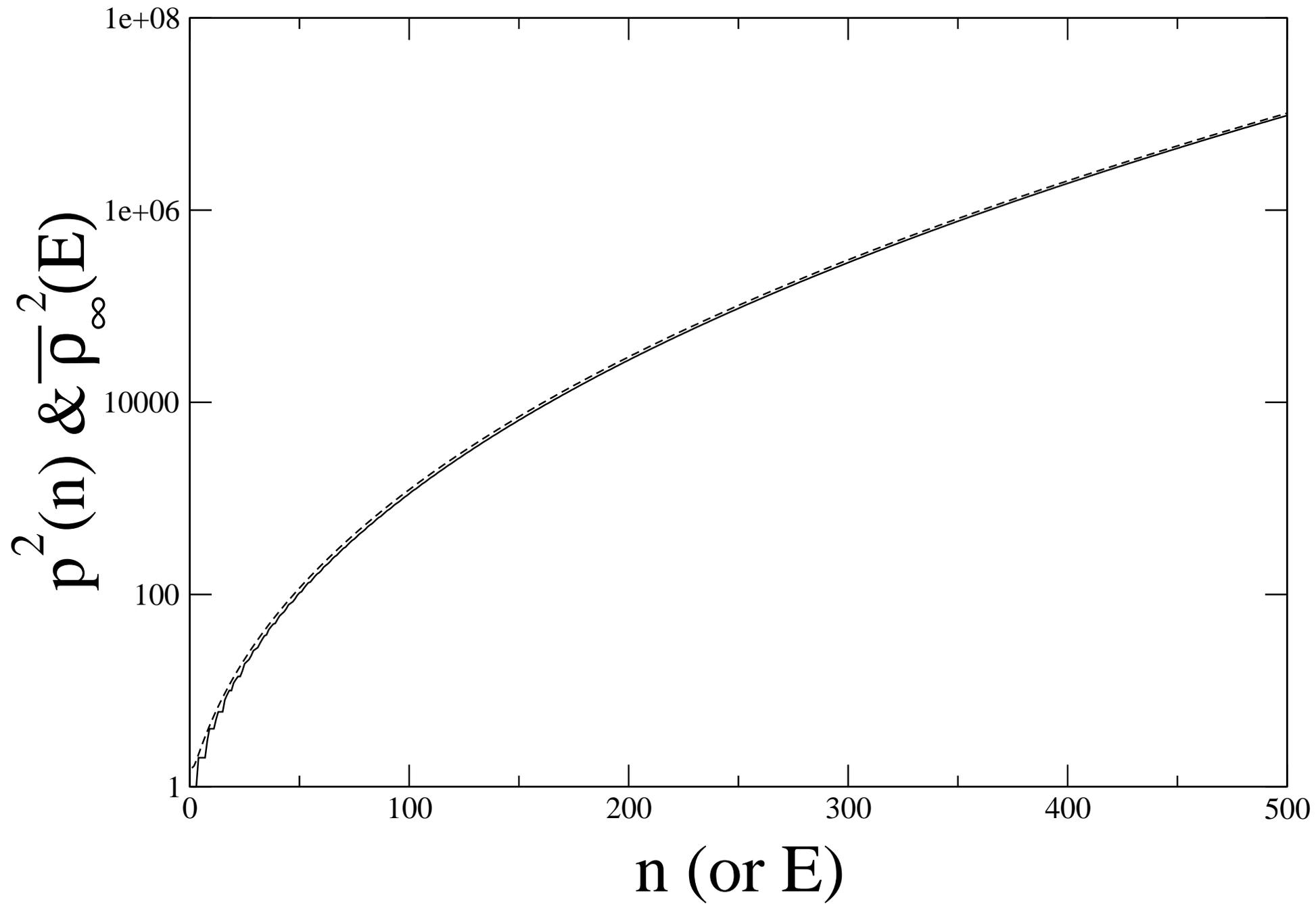
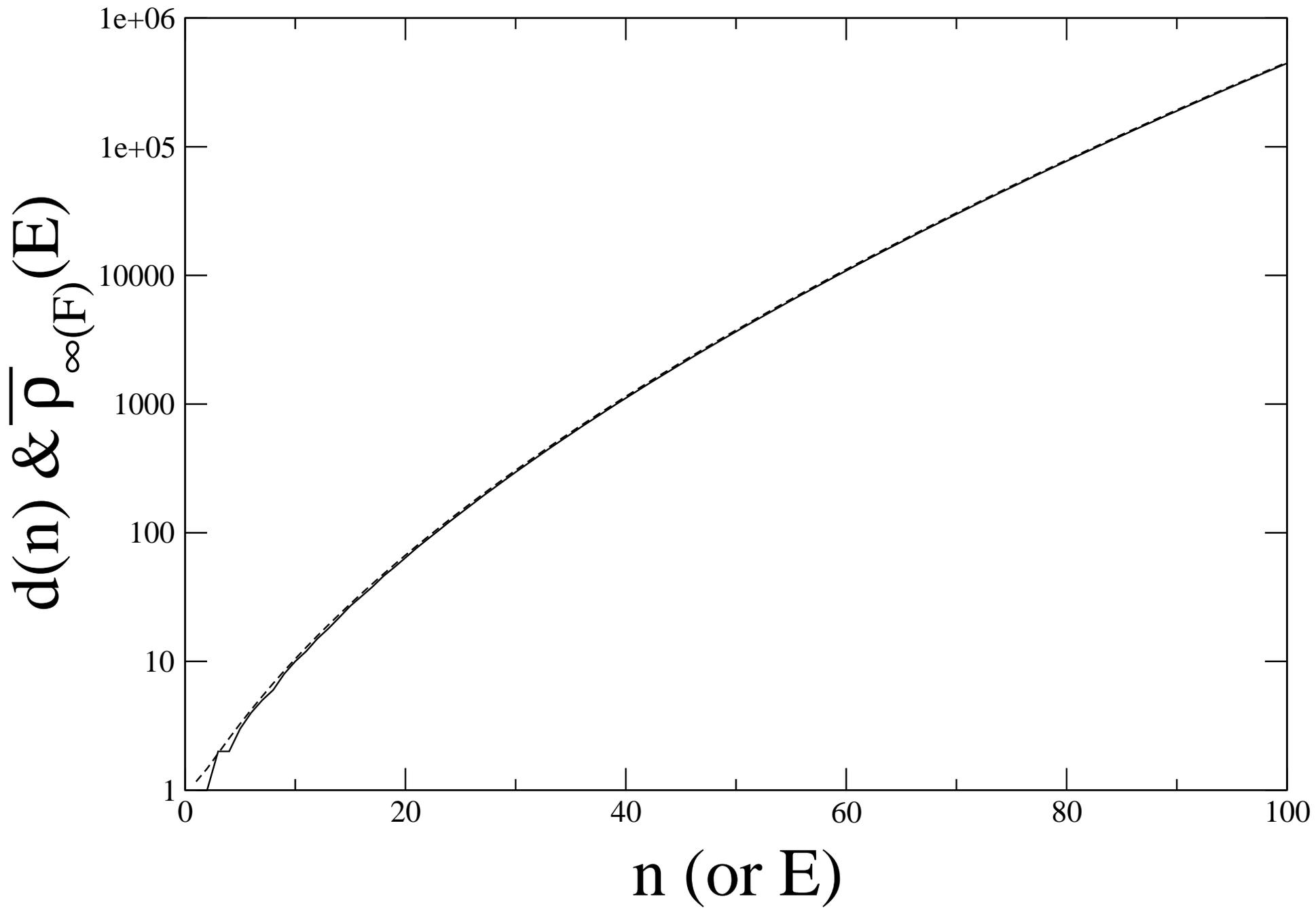
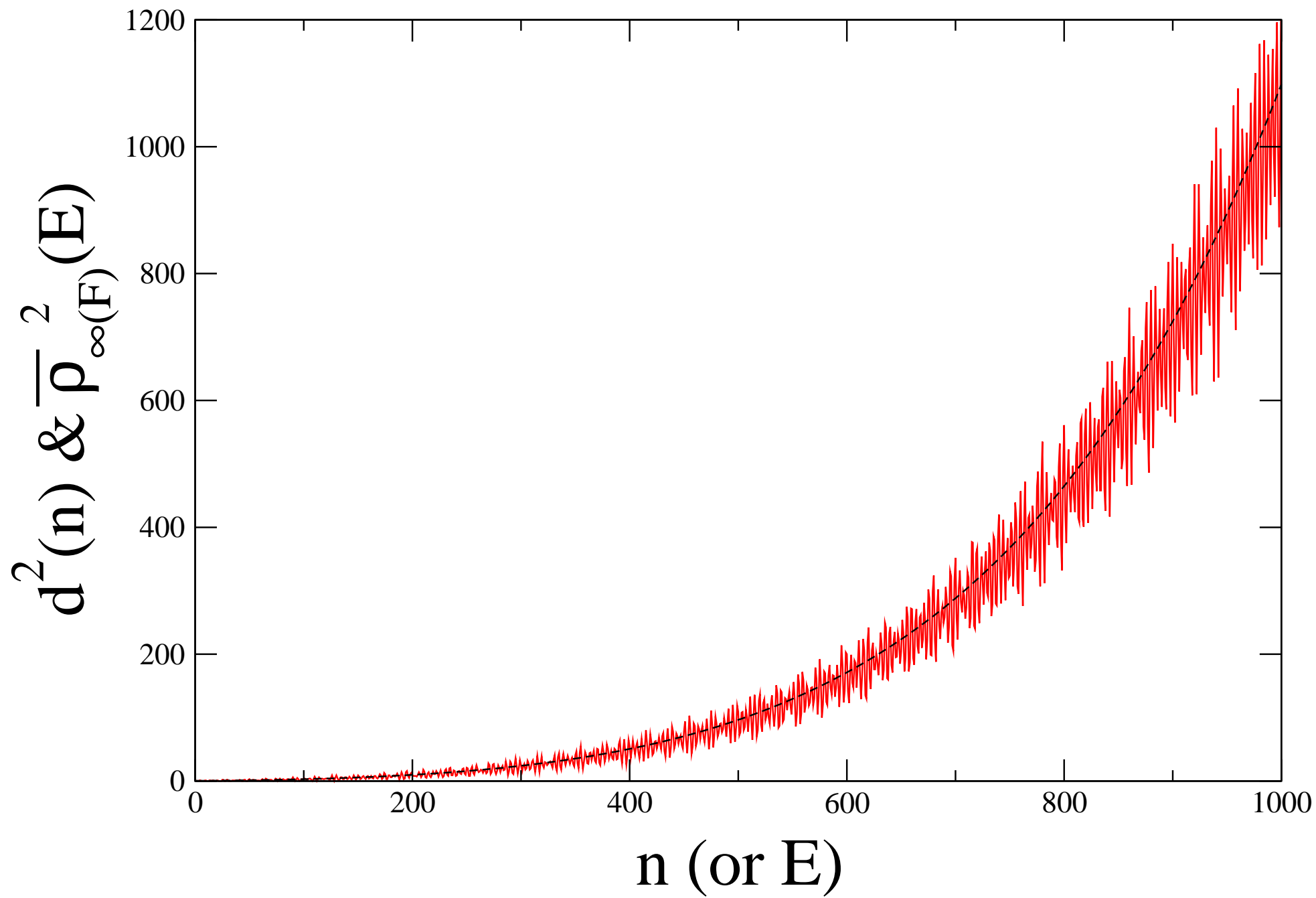
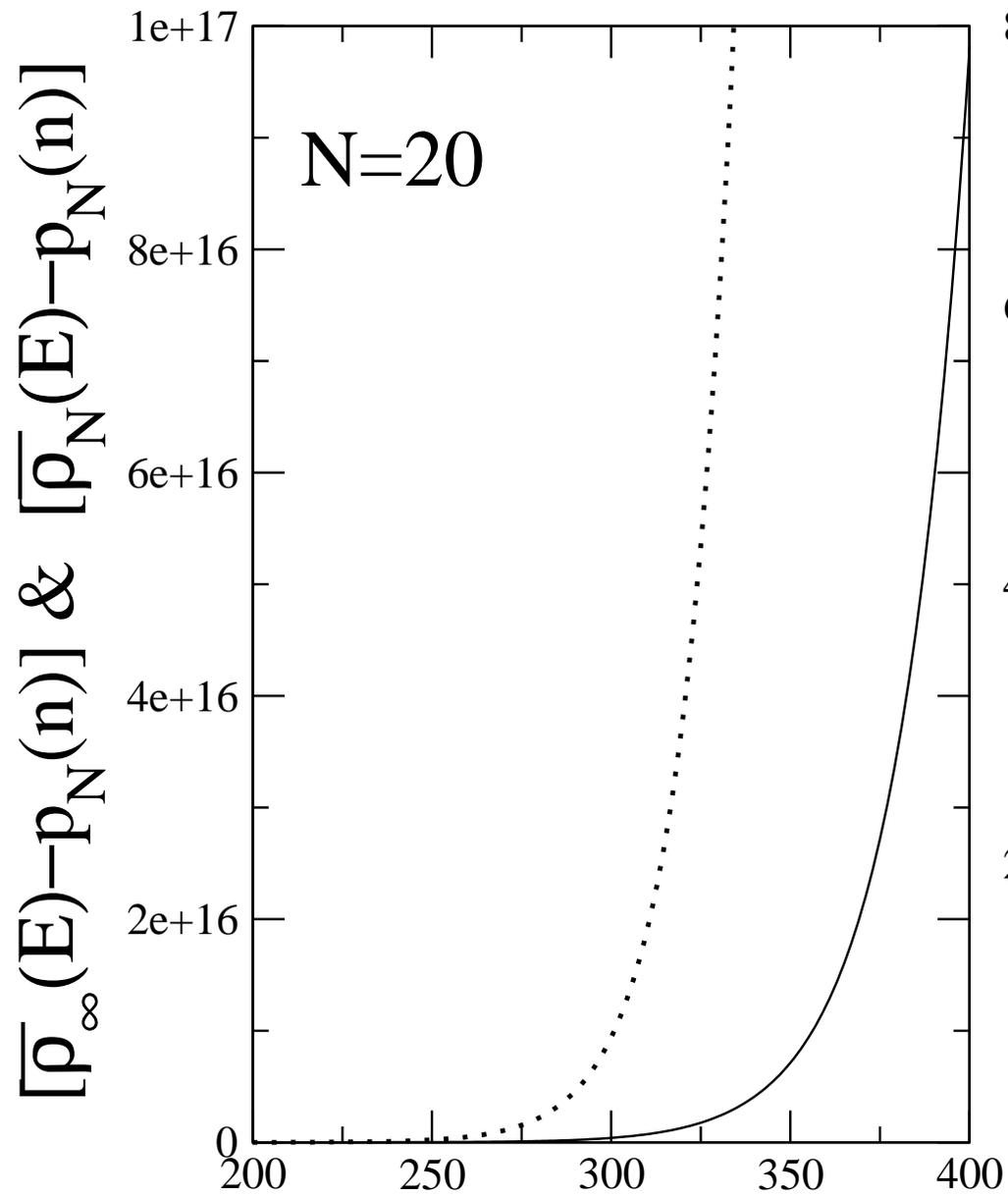
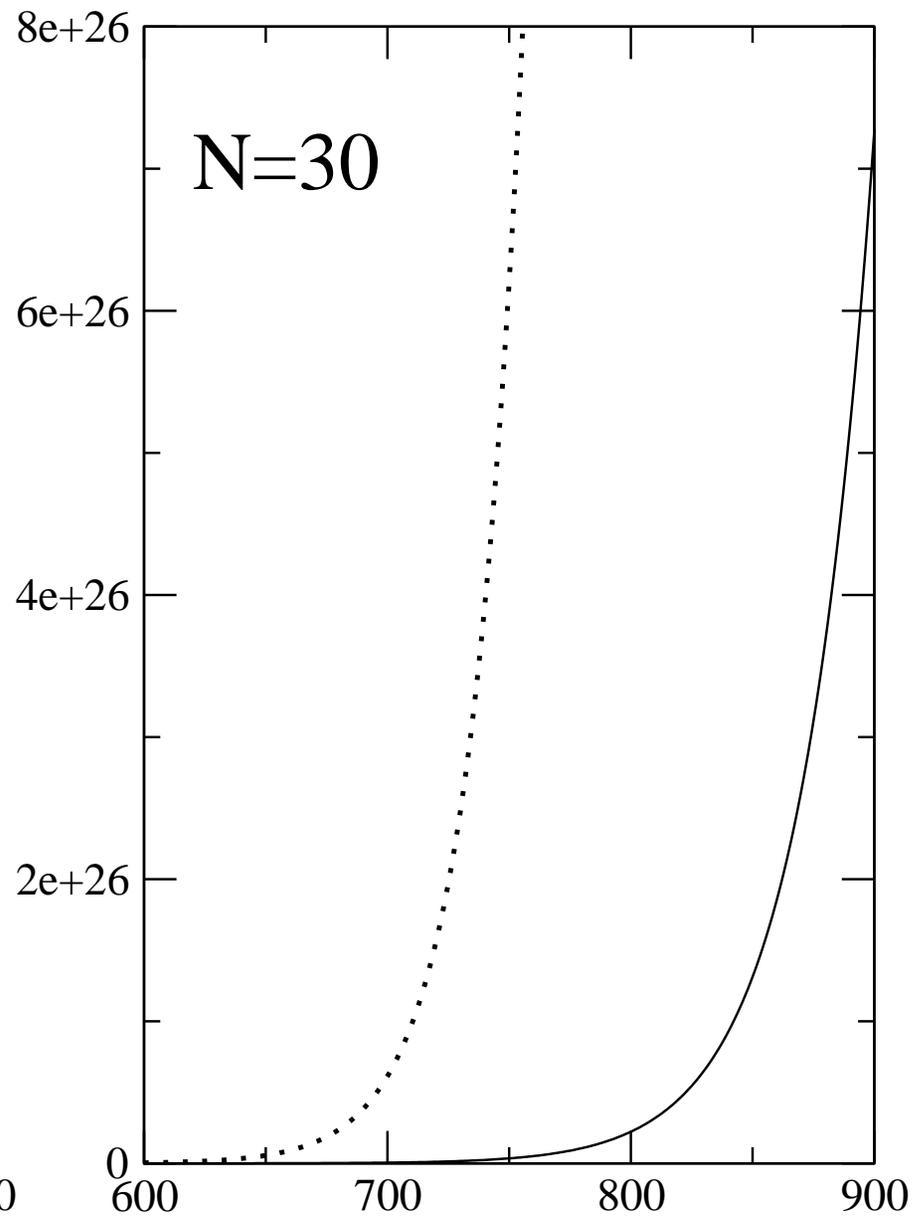
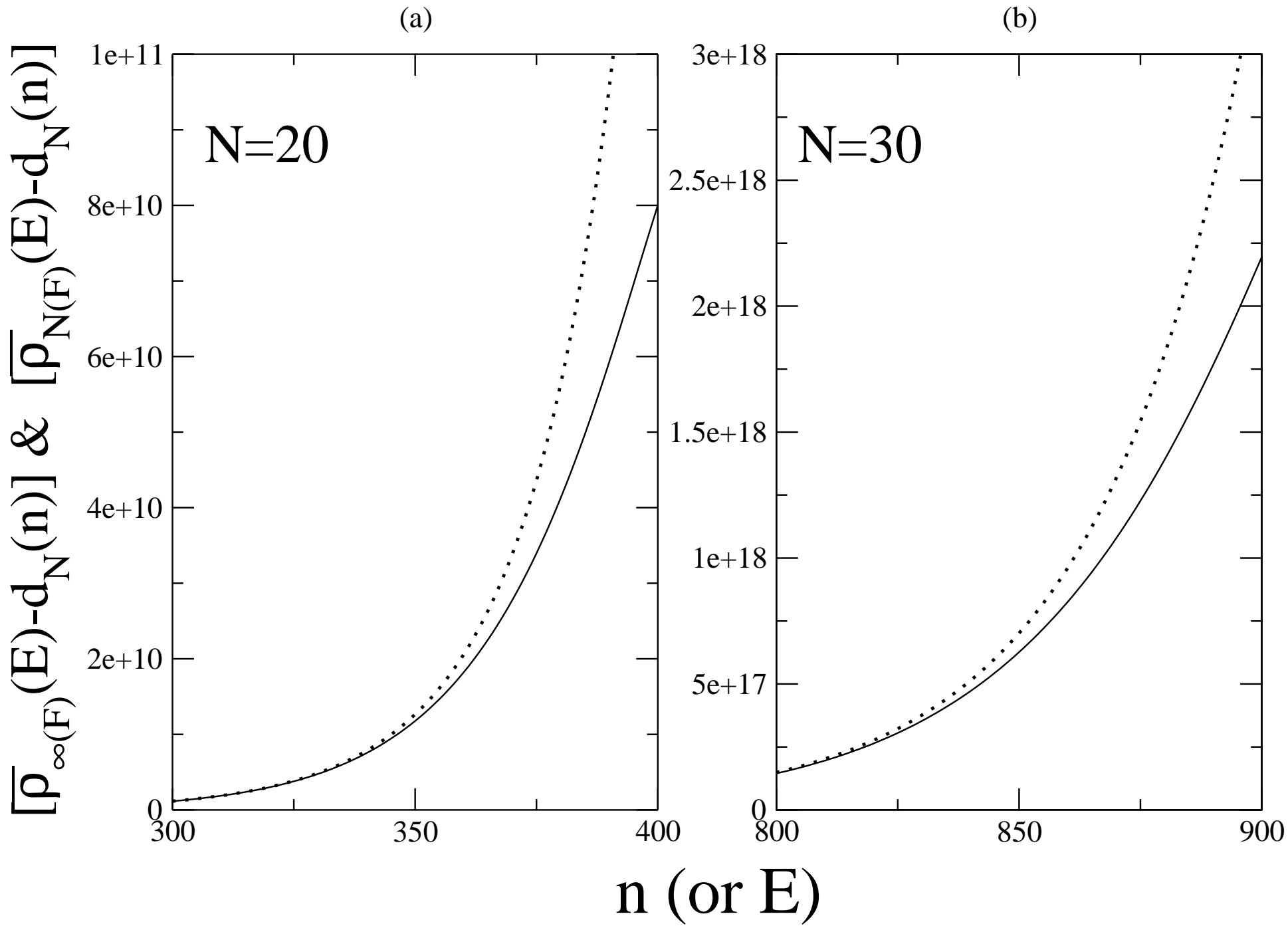
\end{document}